\documentclass{aa}  

\usepackage{txfonts}
\usepackage{graphicx}
\usepackage{natbib}
\usepackage{hyperref}

\newcommand{\SIII}[1]{[\ion{S}{iii}]\ensuremath{~\lambda\ {#1}}}
\newcommand{\SII}[1]{[\ion{S}{ii}]\ensuremath{~\lambda\ {#1}}}
\newcommand{\OIII}[1]{[\ion{O}{iii}]\ensuremath{~\lambda\ {#1}}}

\newcommand{\ArIII}[1]{[\ion{Ar}{iii}]\ensuremath{~\lambda\ {#1}}}
\newcommand{\ArIV}[1]{[\ion{Ar}{iv}]\ensuremath{~\lambda\ {#1}}}

\newcommand{\hbeta}{\ensuremath{H_{\rm \beta}}}

\begin{document} 

\title{Deriving the abundance distribution of HII galaxies using sulphur as tracer: Exploring the high-metallicity end\\}

\titlerunning{Sulphur-based abundance of SDSS HII galaxies}

   \author{M.~Gavil\'{a}n
          \inst{1}          
          A.~I. D\' {i}az\inst{1,2}
          \and
          S.~Zamora\inst{1,2,3}\fnmsep
          }

   \institute{Departamento de F\'{i}sica Te\'{o}rica, Universidad Aut\'{o}noma de Madrid, Madrid 28049, Spain\\            
         \and
             CIAFF, Universidad Autónoma de Madrid, 28049 Madrid, Spain\\ 
        \and   
              Scuola Normale Superiore, Piazza dei Cavalieri 7, I-56126 Pisa, Italy
              }

   \date{}

 
  \abstract
   {}
   {The main objective of this work is to derive the distribution of the the metal content of HII galaxies using sulphur as an abundance tracer. This increases the metallicity range that can safely be reached.}
   {We selected a sample of emission-line galaxies that we extracted from the SDSS-DR16. These objects have a redshift of z$\leq$ 0.04 so that the 
   [\ion{S}{iii}] $\lambda$ 9069 \AA\ emission line and H$_{\beta}$ equivalent widths that are higher than 10 \AA in emission were included, and they are compact in appearance. We used the so-called direct method for objects with the electron-temperature-sensitive [\ion{S}{iii}] $\lambda$ 6312 \AA\ emission line, and an empirical method based on the S$_{23}$ parameter. The last provided an abundance calibration that monotonically increased up to at least the solar value, and can be applied based on the spectral range from 6000 to 9500 \AA\ alone.}
  {We show that the bias that is introduced when the [\ion{O}{iii}] $\lambda$ 4363 \AA\ line is required restricts the sample to objects with an [\ion{O}{iii}] electron temperature higher than  ~10,000K, and their temperature distribution is then rather narrow. For objects with determinations of $t_{\rm e}$[\ion{S}{iii}],  the distribution is flatter and wider, which fits a more realistic scenario better. For these objects, we calculated the ionic abundances of sulphur and their ratios. In all cases, S$^{2+}$ was found to be the dominant ion, with a contribution greatly exceeding that of S$^{+}$. This fact justifies the adoption $t_{\rm e}$[\ion{S}{iii}] as the dominant temperature throughout the nebula, although in 20\% of the objects, an estimated ionization correction factor is required.
  For the objects in the sample that required the detection of the \OIII{4363}~\AA\ line (sample~3) and [\ion{S}{iii}]  $\lambda$ 6312 \AA\,, the distribution abundances as traced directly by oxygen and sulphur appear to be very similar to each other. The median values are 12+log(O/H) = 8.1 and 12+log(S/H) = 6.4, which corresponds to an S/O ratio of log(S/O)= -1.7 that is close to the solar value (-1.5). However, when the restriction for weak temperature-sensitive lines is relaxed,  the abundance distribution is wider and the  median value is 12+log(S/H) = 6.6. When the S/O ratio is assumed to be constant, the median sulphur abundance value found here would imply a median value of the oxygen abundance of 12+log(O/H) = 8.3. }
   {In summary, the abundance distributions traced by sulphur can reach reliable abundances up to the solar value at least and provide a more complete picture of the metallicity distribution of HII galaxies. The method presented here only involves the red part of the spectrum (between 6000 and 9600 $\AA$), and the effect of reddening is weak there. Although the strong nebular [\ion{S}{iii}] lines shift beyond the far red spectral region for high-redshift objects, present-day infrared spectrographs can overcome this difficulty. Observations made with NIRSpec on board the JWST would be able to provide data for objects with redshifts between 0 and 4.24. }


   \maketitle
-------------------------------------------------------------------

\section{Introduction}

   The formation and evolution of galaxies at different cosmological epochs are mainly driven by two processes that are necessarily intertwined: the star formation history, and the metal enrichment. Eventually, the epoch of the first star formation will be revealed based on how far back in time metals in the Universe can be traced. 

The standard cold dark matter model of cosmic evolution currently postulates that the structure in the Universe grew hierarchically. Small objects formed first and increasingly larger systems subsequently formed through mergers. The first galaxies in current simulations form after the first stars, and the first stars cause an initial metal enrichment. The gas has to cool before new stars can be formed, and the forbidden emission lines that are produced by different ions of different elements act as the main coolant agents. Thus, chemical enrichment by the first supernovae is one of the most important processes in the formation of the first galaxies. On the other hand, stars form within galaxies as they assemble, and massive stars evolve fast (~10 Myr) and violently explode as supernovae. This can produce combined gas outflows that are able to deprive galaxies from part of their metals. This pollutes the intergalactic medium.

Nebular abundances are still the preferred and safest method for obtaining present-day (in situ) chemical abundances. Most determinations of the metallicity in galaxies of the Local Group were made through the spectroscopic analysis of nebular gas that is ionized by the UV radiation emitted by hot massive young stars. These determinations are based on radiation theory and atomic physics, which allows us to calculate the elemental abundances of the ions from the measurement of available emission-line intensities by what is usually called the "direct method". However, most of these lines are collisionally excited, and their intensities are measured relative to those of hydrogen-recombination lines. These ratios strongly depend on the electron temperature of the gas. Unfortunately, the diagnostic lines for the gas temperature (so-called auroral lines) are rather weak, and their detection is limited. On the observational side, the method is restricted to objects for which spectra with a sufficiently high signal-to-noise ratio can be obtained so that the weak auroral lines can be measured. On the theoretical side, the line emission via deexcitation facilitates the cooling of the gas, and as the average metallicity of the gas increases, its electron temperature decreases and eventually prevents the observation of the weak auroral lines.

The detection of the weak temperature-sensitive emission lines becomes increasingly easier at redder spectral ranges because the frequency of the involved transitions becomes lower.  This is clearly shown in Fig. 1 of \cite{bre08}, where the auroral  [\ion{O}{iii}] line at 4363 $\AA$ cannot detected by ground-based 8-10m telescopes for electron temperatures below about 8000 K. The corresponding lines of [\ion{N}{ii}] at 5755~$\AA$ and the [\ion{S}{iii}] sulphur line at 6312~$\AA$ are detectable down to 6000 K, however. Therefore, the abundances derived with the direct method are most restrictive for oxygen and are least restrictive for sulphur.

Oxygen has long been used as a tracer of metallicity for various reasons: It is the most abundant element in the interstellar medium (ISM) after hydrogen and helium; it is mainly ejected by massive stars and therefore appears at the early stages of galaxy evolution; and the strong forbidden emission lines of single- and double-ionized atoms can easily be observed in the visible range of the spectrum. Although the lines of thrice-ionized oxygen cannot be observed, the application of an ionization correction factor in star-forming regions is usually not required because this ion accounts for less than 1\% of the total abundance even in extreme emission-line galaxies or active galactic nuclei \citep[AGN;][]{izo06,ber21,dor22}. However, the choice of oxygen as an abundance tracer also presents some drawbacks, as explained below.

The most reliable way to measure the abundance of an element is through the so-called direct method described above, for which it is necessary to know the electron temperature of each ion. However, this temperature is not constant throughout ionized nebulae and shows a negative gradient outward, so that the different oxygen ions O$^{++}$ and O$^{+}$ are formed in different zones with somewhat different temperatures. In order to take this into account, it is necessary to adopt an ionization structure. In a two zone-model, high-excitation ions such as O$^{++}$ are formed in the inner zone, and low-excitation ions such as  O$^{+}$ are formed in the outer zone. Two different electron temperatures, $t_{\rm e}$(O$^+$) and $t_{\rm e}$(O$^{++}$), are therefore needed in order to derive the total oxygen abundance because it can safely be assumed that these two ions alone are present in relevant numbers. However, only one of these temperatures, $t_{\rm e}$(O$^{++}$), is readily available in most cases, and $t_{\rm e}$(O$^+$) is typically derived from photoionization modeling and therefore carries a large uncertainty. This uncertainty in turn introduces large errors in the determination of the O$^+$/H$^+$ abundance. In addition, oxygen is somewhat depleted onto dust grains \citep{izo06}. The uncertainty and depletion are not important for the lowest-metallicity regions, but they might become increasingly relevant at higher abundances and might thus introduce unknown biases in the determination of the metallicity distribution of large object samples. 

The atomic structure of sulphur is similar to that of oxygen. Sulphur is also produced by massive stars and is hence thought to appear in the ISM at the same time, but the emissivity of its [\ion{S}{ii}] and [\ion{S}{iii}] lines depends more weakly on electron temperature. S$^{++}$ is formed in what is called the intermediate excitation zone, which overlaps the two zones described above and appears to be wide enough for the [\ion{S}{iii}] line electron temperature to be representative of the whole nebula, as shown by photoionization models  \citep{vit18}. This allows the approximation of  $t_{\rm e}$([\ion{S}{iii}]) $\sim$ $t_{\rm e}$([\ion{S}{ii}]), $t_{\rm e}$([\ion{O}{ii}]), $t_{\rm e}$([\ion{N}{ii}]). 
Furthermore, $t_{\rm e}$([\ion{S}{iii}]) can be derived from the quotient between the nebular and auroral [\ion{S}{iii}] lines at $\lambda\lambda$ 9069, 9532 \AA\ and $\lambda$ 6312 \AA\, respectively, all of which lie in the red part of the spectrum and can easily be detected at metallicities up to the solar value. Neither the auroral nor the nebular oxygen emission lines lie at this value  \citep[see, for example][]{dia07}. This presents two advantages: First, the contribution of these lines to the cooling is more effective at low temperatures, which allows us to reach higher abundances of up to solar at least with the direct method. Second, this spectral range presents fewer metal absorption lines and is less affected by extinction, which reduces the reddening effects. On the other hand, at the highest temperatures, that is, in low-metallicity nebulae, there is a non-negligible fraction of S$^{3+}$ (with a forbidden emission line in the infrared (10.4 $\mu$m)), and as a consequence, the ionization correction factors (ICF) need to be estimated.


The limiting abundance range in which the oxygen abundances in HII galaxies can be derived in summary precludes us from knowing the entire metallicity distribution of these objects. Although the [\ion{S}{iii}] lines are included in Sloan Digital Sky Survey (SDSS) spectra for objects up to redshift $\sim$ 0.014, they are rarely used. However, they provide an excellent opportunity to explore the high-metallicity end of emission-line galaxies with a homogeneous sample, and we  can thus obtain a more realistic abundance distribution for HII galaxies. This is the purpose of the present work. We have selected star-forming emission-line galaxies that appeared to be compact from the SDSS 16th Data Release with data on the \SIII{9069} \AA\ line. For these objects, we measured the fluxes of the main emission lines and derived the sulphur abundance with the aim of determining the actual and complete metallicity distribution of the sample. In Sect.~\ref{data} we describe the selection of the sample, and in Sect.~\ref{metodo_medida} we detail the way in which the line fluxes were measured. In Sect.~\ref{resultados} we explain the abundance derivation methods, and we discuss our results in Sect.~\ref{discusion}. Finally, our conclusions are set out in Sect.~\ref{conclusiones}.

\section{Sample selection}
\label{data}
\subsection{SDSS object selection}
The data analyzed in this work were extracted from the DR16 release \citep{sd16r}. A query to this database was made according to the following criteria: 
\begin{itemize}
\item Redshift, \textit{z}, $<$ 0.014. 
\item Equivalent  width of the \hbeta\ emission line,  EW(\hbeta) $>$ 10\AA . 
\item Signal-to-noise ratio (S/N) at $\hbeta$  and at \OIII{5007} $>$ 2.
\item Type: GALAXY.
\end{itemize}
 
Objects with a redshift lower than 0.014 were selected in order to include the  \SIII{9069} \AA\ emission line within the spectral range. The constraints we set up for the  $\hbeta$ equivalent width and the \OIII{5007} \AA\ line intensity also allowed us to select spectra with strong emission lines and a low continuum level, for which intense star formation is assumed to be the main ionization mechanism.

\begin{figure}
\includegraphics[height=7cm, width=9cm]{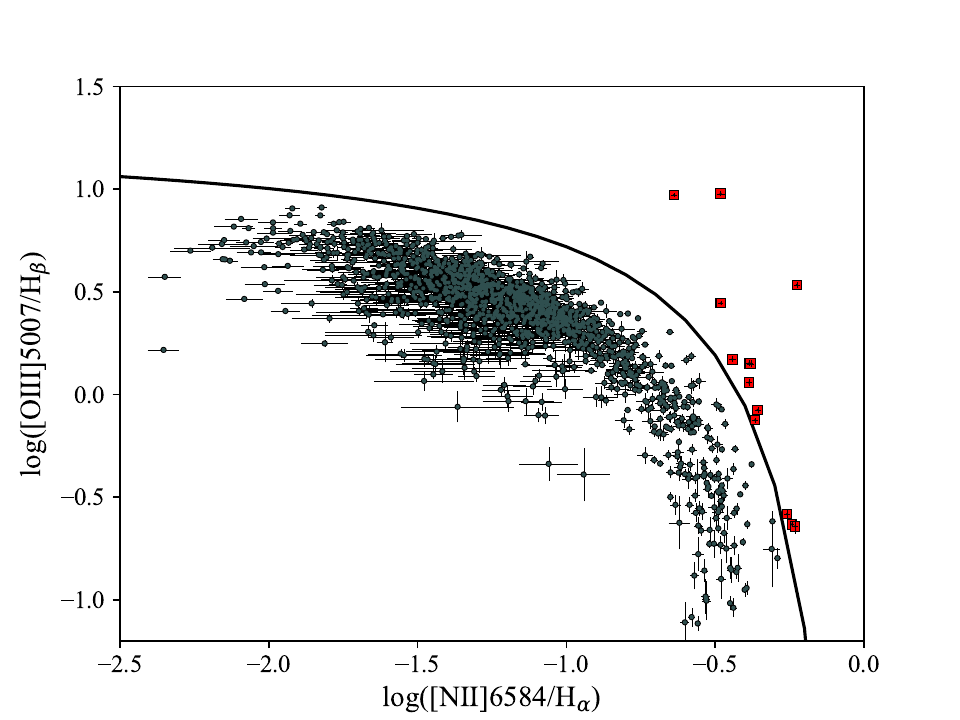}
\caption{Classical BPT diagram. The solid black line was described in \citet{kew06}. The fluxes and errors are those provided by the SDSS database. The objects marked with red squares were removed from our analysis.}
\label{BPT}
\end{figure}

The flux of the \SIII{9069} \AA\ line is not included in the SDSS data archive server list of measured fluxes. We therefore developed our own tool to measure it (see Sect. ~\ref{metodo_medida}). With this tool, we selected only objects from the initial list whose (S/N) in the \SIII{9069} , \SII{6717}, and \SII{6731} \AA\ emission lines was higher than 2. We discarded the rest. In order to identify the ionization source, a Baldwin, Phillips \& Terlevich (BPT)  diagram \citep{BPT} was produced  using the fluxes provided by the SDSS database (Fig.~\ref{BPT}). This further reduced the sample to objects that were consistent with a thermal origin for their ionization.  We also inspected the images of all objects visually to select only objects that were compact enough to be considered HII galaxies. This collection is called sample~1 and is composed of 439 objects.

Out of this sample, we chose objects with a detected auroral sulphur line \SIII{6312} \AA\ in their spectra. This allowed us to obtain the value of the electron temperature, $t_{\rm e}[\ion{S}{iii}]$, and based on this, to determine the sulphur abundances using the so-called direct method. The result was a list of 224 objects that we call sample~2.

Finally, we have composed sample~3, which is more restrictive and contains only the elements from sample 2 that show measurable \OIII{4363} \AA\ and ${[\ion{O}{ii}]}~\lambda\lambda\ {7322,7330}$ \AA\ emission lines for which oxygen abundances can also be derived by the direct method to be able to simultaneously compare the abundance distributions of sulphur and oxygen. These conditions were met for 148 objects. Only 35 objects of this latter sample allowed the detection of the \ArIII{7135} and \ArIV{4740} \AA\ emission lines. We used these to obtain the sulphur ionization correction factor (ICF), as we explain below (see Sect. \ref{S-ICF}).

\section{Measuring the emission-line intensity}
\label{metodo_medida}
 
For consistency, we applied our flux-measuring tool described below to all the emission lines we used. The first step of the procedure consists of subtracting the stellar continuum contribution with a stellar population synthesis model. For this task, we chose the STARLIGHT code \citep{starlight1, starlight2} with the spectral base from \cite{BC03}.
Then, we measured the line fluxes using a single-Gaussian fitting plus a linear term as follows: (i) A global continuum was adjusted after we masked all the emission lines in the spectrum with a width of $\pm$8 \AA\ around the central wavelength of the line. Subsequently, a second-order polynomial was fit, and the statistical dispersion of the continuum, $\Sigma_c$, was calculated. (ii) The selected emission lines were measured using a local continuum set at the central wavelength of the line, $F_{c,\lambda} \pm  \Sigma_c$, and fitting a Gaussian with a width $\sigma_l$  according to the continuum level that was compatible with the dispersion of the global spectrum, according to the expression  

\begin{equation}
F(\lambda) = A_l \cdot \sigma_l \sqrt{2\pi}.
\end{equation}

\noindent Finally, the errors in the measured fluxes were calculated using the expression given in \cite{gon94}, 

\begin{equation}
\sigma_{l} = \sigma_{cont} N^{1/2} [1+\frac{EW}{N\Delta}]^{1/2},
\end{equation}

\noindent where $\sigma_{cont}$ is the standard deviation in the continuum near the line, N is the number of pixels, EW is the equivalent width of the line, and $\Delta$ is the spectral dispersion in $\AA$ per pixel.

\begin{figure}
\includegraphics[width=0.5\textwidth]{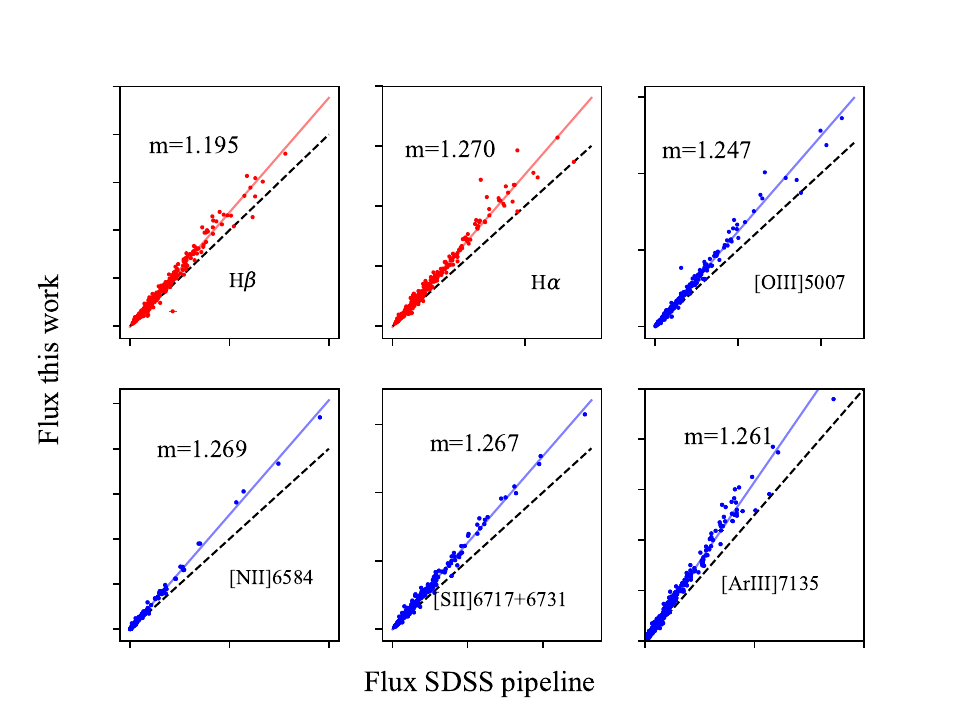}
\caption{Comparison between the fluxes calculated in this work and those published on the SDSS website for some emission lines as indicated in each panel. The dashed black line represents the 1:1 relation, and the solid line is the linear fit to the  data. The slope of the fit, \textit{m}, is given in each panel.}
\label{comparacion}
\end{figure}

Figure~\ref{comparacion} shows the values of the fluxes obtained with our measurement method versus those provided by the SDSS data archive server. Although the correlation is rather good, our values are slightly higher in a very similar proportion in all cases. This suggests a systematic effect that is probably related to  the calculation and subtraction of the continuum flux \citep{nag06}. The slope for the H$_{\alpha}$ line is slightly steeper than for the rest of lines means that the correction for reddening is also slightly larger in our case, although this is not expected to affect the line ratios.

The final intensities were obtained after correcting the observed fluxes for reddening. To do this, we used the ratio of the observed and theoretical Balmer line ratios \citep[see, for example][]{hag08}, 

\begin{equation}
    log\left[\frac{I(\lambda)}{I(H_{\beta})}\right] = log \left[\frac{F(\lambda)}{F(H_{\beta})} \right] + c(H_{\beta})  f(\lambda).
\end{equation}

The final corrected line-intensity values were normalized to H$_{\beta}$ = 100. As an example, Table~1 shows these values for the first six objects of the sample along with their uncertainties. The full table is available at the CDS.

\begin{table*}
\label{tabla_intensidades}
\centering

\caption{Reddening-corrected emission-line intensities.}
\scriptsize
\begin{tabular}{|l|c|c|c|c|c|c|c|c|}
\hline
Object& [OIII]$\lambda$ 4363 \AA & [ArIV]$\lambda$ 4740 \AA & H$\beta$ & [OIII]$\lambda$ 4959 \AA & [OIII]$\lambda$ 5007 \AA &[SIII]$\lambda$ 6312 \AA & H$\alpha$ &[SII]$\lambda$ 6717 \AA ...\\
\hline
\href{http://skyserver.sdss.org/dr14/en/tools/explore/summary.aspx?plate=0266&mjd=51602&fiber=0089}{{\underline{0266-51602-0089}}} & 12.32 $\pm$ 0.72 & 1.01 $\pm$ 0.31 & 100.00 $\pm$ 0.56 & 188.29 $\pm$ 1.70 & 548.58 $\pm$ 3.81 & 1.56 $\pm$ 0.09 & 286.68 $\pm$ 1.70 & 7.13 $\pm$ 0.10 \\
\href{http://skyserver.sdss.org/dr14/en/tools/explore/summary.aspx?plate=0266&mjd=51630&fiber=0100}{{\underline{0266-51630-0100}}} & 13.19 $\pm$ 0.77 & 1.00 $\pm$ 0.32 & 100.00 $\pm$ 0.56 & 184.36 $\pm$ 1.72 & 541.99 $\pm$ 3.81 & 1.30 $\pm$ 0.08 & 287.12 $\pm$ 1.69 & 6.05 $\pm$ 0.09 \\
\href{http://skyserver.sdss.org/dr14/en/tools/explore/summary.aspx?plate=0284&mjd=51662&fiber=0164}{{\underline{0284-51662-0164}}} & 11.13 $\pm$ 0.90 & 0.88 $\pm$ 0.49 & 100.00 $\pm$ 0.95 & 178.36 $\pm$ 2.18 & 532.76 $\pm$ 5.58 & 1.42 $\pm$ 0.27 & 286.39 $\pm$ 2.85 & 10.33 $\pm$ 0.23 \\
\href{http://skyserver.sdss.org/dr14/en/tools/explore/summary.aspx?plate=0284&mjd=51943&fiber=0170}{{\underline{0284-51943-0170}}} & 11.14 $\pm$ 0.74 & 0.47 $\pm$ 0.37 & 100.00 $\pm$ 0.65 & 181.37 $\pm$ 1.65 & 546.96 $\pm$ 4.02 & 1.50 $\pm$ 0.15 & 286.26 $\pm$ 2.03 & 10.15 $\pm$ 0.19 \\
\href{http://skyserver.sdss.org/dr14/en/tools/explore/summary.aspx?plate=0287&mjd=52023&fiber=0466}{{\underline{0287-52023-0466}}} & 6.50 $\pm$ 1.08 & ...   & 100.00 $\pm$ 0.85 & 187.12 $\pm$ 1.86 & 557.70 $\pm$ 4.92 & 1.48 $\pm$ 0.23 & 286.27 $\pm$ 2.51 & 16.07 $\pm$ 0.27 \\
\href{http://skyserver.sdss.org/dr14/en/tools/explore/summary.aspx?plate=0297&mjd=51959&fiber=0019}{{\underline{0297-51959-0019}}} & 7.82 $\pm$ 0.54 & ...   & 100.00 $\pm$ 0.32 & 134.99 $\pm$ 0.53 & 410.90 $\pm$ 1.33 & 2.56 $\pm$ 0.16 & 286.55 $\pm$ 0.92 & 30.63 $\pm$ 0.16 \\

\hline
Object&...[SII]$\lambda$ 6731 \AA &[ArIII]$\lambda$ 7135 \AA &[OII]$\lambda$ 7322 \AA &[OII]$\lambda$ 7330 \AA &[SIII]$\lambda$ 9069 \AA &Flux($H\beta$)10$^{-17}$ $erg/s/cm^{2}$ &EW(H$\beta$)&c(H$\beta$) \\

\hline
\href{http://skyserver.sdss.org/dr14/en/tools/explore/summary.aspx?plate=0266&mjd=51602&fiber=0089}{{\underline{0266-51602-0089}}} & 5.38 $\pm$ 0.09 & 5.17 $\pm$ 0.54 & 1.51 $\pm$ 0.20 & 1.14 $\pm$ 0.19 & 10.28 $\pm$ 0.18 & 6764.57 $\pm$ 38.0 & 460.21 & 0.34 $\pm$ 0.02 \\
\href{http://skyserver.sdss.org/dr14/en/tools/explore/summary.aspx?plate=0266&mjd=51630&fiber=0100}{{\underline{0266-51630-0100}}} & 4.68 $\pm$ 0.08 & 4.02 $\pm$ 0.41 & 1.23 $\pm$ 0.16 & 0.93 $\pm$ 0.16 & 8.01 $\pm$ 0.12 & 6800.33 $\pm$ 38.1 & 445.46 & 0.57 $\pm$ 0.02 \\
\href{http://skyserver.sdss.org/dr14/en/tools/explore/summary.aspx?plate=0284&mjd=51662&fiber=0164}{{\underline{0284-51662-0164}}} & 7.91 $\pm$ 0.21 & 7.06 $\pm$ 0.45 & 1.82 $\pm$ 0.25 & 1.54 $\pm$ 0.26 & 11.00 $\pm$ 0.68 & 1443.07 $\pm$ 13.7 & 178.33 & 0.20 $\pm$ 0.03 \\
\href{http://skyserver.sdss.org/dr14/en/tools/explore/summary.aspx?plate=0284&mjd=51943&fiber=0170}{{\underline{0284-51943-0170}}} & 7.98 $\pm$ 0.18 & 6.66 $\pm$ 0.45 & 1.84 $\pm$ 0.27 & 1.08 $\pm$ 0.25 & 12.28 $\pm$ 0.78 & 1598.44 $\pm$ 10.4 & 194.34 & 0.13 $\pm$ 0.02 \\
\href{http://skyserver.sdss.org/dr14/en/tools/explore/summary.aspx?plate=0287&mjd=52023&fiber=0466}{{\underline{0287-52023-0466}}} & 11.29 $\pm$ 0.24 & 6.93  0.29 & 2.54 $\pm$ 0.31 & 1.78 $\pm$ 0.32 & 16.58 $\pm$ 0.50 & 1291.44 $\pm$ 10.9 & 40.94 & 0.14 $\pm$ 0.03 \\
\href{http://skyserver.sdss.org/dr14/en/tools/explore/summary.aspx?plate=0297&mjd=51959&fiber=0019}{{\underline{0297-51959-0019}}} & 21.80 $\pm$ 0.15 & 7.60  0.18 & 2.73 $\pm$ 0.18 & 1.50 $\pm$ 0.18 & 14.19 $\pm$ 0.26 & 373.02 $\pm$ 1.2 & 40.73 & 0.28 $\pm$ 0.01 \\

\hline
\end{tabular}

\tablefoot{Reddening-corrected emission-line intensities. We only list the data for the first six objects as an example. The table is too wide and was split into two for clarity. The full table is available at the CDS.}
\end{table*}
\normalsize

\section{Results}
\label{resultados}

\subsection{Direct sulphur abundances}

To derive the elemental abundances, the physical conditions of the ionized gas need to be known, mainly the electron density and temperature. The first can be estimated from the [\ion{S}{ii}]~$\lambda $~6717 \AA\/$\lambda$~6731 \AA\ emission-line ratio. Because our objects are HII galaxies, the electron density of the emitting gas was found to be lower than 500 ~cm$^{-3}$ in most cases \citep{izo06}, which is well below the critical value for collisional deexcitation. For these objects, an uncertainty in the density of about 200 ~cm$^{-3}$ would lead to an error in the final derived abundances smaller than 1\% \citep{ver02}. Hence, we assumed a mean value of 100  ~cm$^{-3}$ for all the analyzed objects. The second condition can be derived from the ratio of the nebular to auroral [\ion{S}{iii}] emission lines at $\lambda\lambda$~9069,9532 \AA\ and $\lambda$~6312 \AA.

Following \cite[hereinafter DZ22]{dia22}, to derive $t_{\rm e}$[\ion{S}{iii}], we used the relation

\begin{equation}
t_{e}[\ion{S}{iii}] = 0.5597+1.808\cdot 10^{-4} R_{S3}+\frac{22.6575}{R_{S3}},
\label{t_SIII}
\end{equation}

\noindent where
\[ R_{S3} = \frac{I(\SIII{9069}) + I(\SIII{9532} )}{I(\SIII{6312})} \].
Since the analyzed spectra do not reach wavelengths longer than 9200 $\AA$, we used the theoretical ratio of the [\ion{S}{iii}] $\lambda$~9532 \AA\ and $\lambda$~9069 \AA\ line intensities, which were taken to be 2.44, to take the contribution of the [\ion{S}{iii}] $\lambda$ 9532 \AA\
line into account.

The method for deriving the sulphur abundance was thoroughly explained in Díaz and Zamora (2022) for extragalactic HII regions and HII galaxies. According to the ionization structure of this type of object, as discussed in \cite{vit18}, the S$^{+}$ and S$^{2+}$ ions are both present in the intermediate-ionization zone of the nebula, and a common electron temperature can be considered in this region, that is, $t_{\rm e}[\ion{S}{iii}]$ = $t_{\rm e}[\ion{S}{ii}]$. In this scheme, the nebula can be divided into two temperature zones: a high-temperature zone, represented by $t_{\rm e}$([\ion{O}{iii}]), in which the [\ion{O}{iii}], [\ion{Ar}{iv}], and [\ion{S}{iv}] lines originate, and an intermediate- to low-temperature zone, represented by $t_{\rm e}$([\ion{S}{iii}]), in which the [\ion{S}{iii}], [\ion{Ar}{iii}], [\ion{S}{ii}], [\ion{O}{ii}], and [\ion{N}{ii}] are produced.

Because no [\ion{S}{iv}] lines lie in the visible part of the spectrum, the derivation of the sulphur abundance was reduced to considering a single zone with only one electron temperature, $t_{\rm e}$([\ion{S}{iii}]). The contribution by S$^{3+}$ needs to be obtained by other means.

Ionic abundances were calculated by fitting the abundance relation calculated using PyNeb \citep{pyneb} with the atomic data reported by \cite{fro04} and \cite{kis09}. We assumed the electron density to be equal to 100~cm$^{-3}$ in all cases, as mentioned above. The resulting expressions are listed below.

\begin{eqnarray}
12+log\left(\frac{S^{+}}{H^{+}}\right)=log\left(\frac{I(6717+6731)}{H_{\beta}}\right)+5.516 + \nonumber\\
+\frac{0.884}{t_{e}}-0.480log(t_{e}),
\label{eq_1}
\end{eqnarray}

\begin{eqnarray}
12+log\left(\frac{S^{2+}}{H^{+}}\right)=log\left(\frac{I(9069+9532)}{H_{\beta}}\right)+6.059 + \nonumber\\
+\frac{0.608}{t_{e}}-0.706log(t_{e}),
\end{eqnarray}

\noindent where $t_{\rm e}$ denotes the electron temperature, $t_{\rm e}$([\ion{S}{iii}]), in units of 10$^{4}$K.


\subsection{Direct oxygen abundances}
\label{oxigeno}
Although the objective of this work is to use sulphur as metallicity tracer, we also calculated the abundance of oxygen in order to be able to estimate the ionization correction factor.

Similar to the case for sulphur, the [\ion{O}{iii}] electron temperature can be found as \citep{hag08}

\begin{equation}
t_{\rm e}[\ion{O}{iii}] = 0.8254-2.45\cdot 10^{-4} R_{O3}+\frac{47.77}{R_{O3}},
\end{equation}

where
\begin{equation}
R_{O3} = \frac{\OIII{4959} + \OIII{5007}}{\OIII{4363}}.
\end{equation}

In the same way as for sulphur, we fit the ionic abundances using Pyneb,

\begin{eqnarray}
12+log\left(\frac{O^{2+}}{H^{+}}\right)=log\left(\frac{I(4959+5007)}{H_{\beta}}\right)+6.249 + \nonumber\\
+\frac{1.184}{t_{e}}-0.708log(t_{e}),
\end{eqnarray}

\noindent where $t_{\rm e}$ is the [\ion{O}{iii}] electron temperature in units of 10$^{4}$K, and the electron density was assumed to be 100 cm$^{-3}$.

As mentioned in Sect. \ref{resultados}, we assumed only two temperature zones, and $t_{\rm e}$[\ion{O}{ii}] can be taken as equal to $t_{\rm e}$[\ion{S}{iii}]. The electron temperatures for oxygen are shown in Table~2, which is only provided in electronic form at the CDS.

To calculate the ionic abundances of O$^{+}$/H$^{+}$, the [\ion{O}{ii}]  $\lambda \lambda$~7322,7330 \AA\ lines were used because the spectra we studied here do not include the blue [\ion{O}{ii}] lines at $\lambda\lambda$~ 3727,29 \AA . The red lines can be affected by recombination, and it is necessary to subtract this component from the total flux. For this task, we used Eq.~(2) from \cite{liu00} ,

\begin{equation}
\frac{I_{rec}(7322+7330)}{H_{\beta}} = 9.36  t^{0.44} \times \frac{O^{++}}{H^{+}}.
\end{equation}

The mean value of this contribution to the total flux in our spectra is about 4\%, and it does not exceed 8\% throughout. The O$^+$/H$^+$ ionic ratio was calculated as 

\begin{eqnarray}
12+log\left(\frac{O^{+}}{H^{+}}\right)=log\left(\frac{I(7322+7330)}{H_{\beta}}\right)+6.952+ \nonumber\\
+\frac{2.433}{t_{e}}-0.571log(t_{e}),
\label{o_mas}
\end{eqnarray}

\noindent where I(7322+7330) is the collisional contribution of the total flux, and $t_{\rm e}$ is equal to $t_{\rm e}$[\ion{S}{iii}].

Although the objects of our sample were classified as HII galaxies and are thought to have a low metallicity and high electron temperature,  the HeII emission line at $\lambda$~4686 \AA\ was not detected. This indicates that the contribution of $O^{3+}$ is negligible, and thus, no ICF is needed in this case. The total O/H abundance can then be calculated as

\begin{equation}
\frac{O}{H} = \frac{O^{+}+O^{++}}{H^{+}}.
\end{equation}

The oxygen ionic and total abundances for sample~3 objects are shown in Table~2, which is only provided in electronic form at the CDS.

\subsection{Sulphur ionization correction factor}
\label{S-ICF}

In some of the objects we studied, the ionizing temperature is sufficiently high for sulphur to be ionized three times. A certain contribution to the total S abundance by $S^{3+}$ is therefore expected and had to be taken into account. No [\ion{S}{iv}] lines lie within the optical spectral range, and this can therefore be done by calculating an ionization correction factor (ICF), which is defined as

\begin{equation}
\frac{S}{H} = ICF\cdot \frac{S^{+} + S^{2+}}{H^{+}}.
\end{equation}

\begin{figure}
\begin{center}
\includegraphics[width=0.4\textwidth]{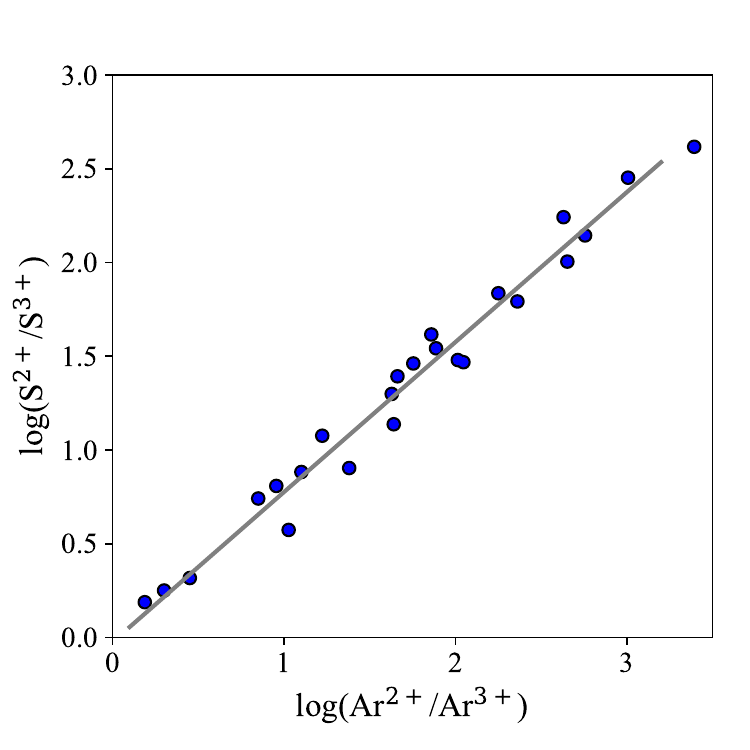}
\caption{Relation of two consecutive ionization stages of argon and sulphur derived using a grid of Cloudy photoionization models as described in the text.} 
\label{ajuste_Argon}
\end{center}
\end{figure}

We used the relation of two consecutive ionization stages of argon and sulphur \citep{vit18} derived using a grid of Cloudy photoionization models \citep{cloudy} to calculate the S$^{3+}$/S$^{2+}$ as a function of Ar$^{3+}$/Ar$^{2+}$. This can be derived from observations for the objects in sample 3, which include the lines of [\ion{Ar}{iii}]$\lambda$~7135 \AA\ and [\ion{Ar}{iv}]$\lambda$~4740\AA .
All models were calculated under the assumption of spherical symmetry and a constant electron density of 100 particle cm~$^{-3}$. Considering that the distance to the emitting gas shell is larger than its thickness, we assumed a plane-parallel geometry approximation. The grid of models was characterized by logarithmic ionization parameters between -2 and -3, ionizing star clusters with ages between 2 and 5 Ma \citep[PopStar models][]{mol09}, and metallicities between Z$_{\odot}$/20 and Z$_{\odot}$/2. The resulting fit is shown in Fig.~\ref{ajuste_Argon}.

A linear fit of the data gives 

\begin{equation}
log\left(\frac{S^{2+}}{S^{3+}}\right) = (0.80 \pm 0.03) log\left(\frac{Ar^{2+}}{Ar^{3+}}\right) + (-0.02 \pm 0.05).
\label{Ar_ICF}
\end{equation}

The ionic abundances of argon, Ar$^{2+}$/Ar$^{3+}$, were calculated using PyNeb as 

\begin{eqnarray}
12+log\left(\frac{Ar^{2+}}{H^{+}}\right)=log\left(\frac{I(7135)}{H_{\beta}}\right)+6.145 + \nonumber\\
+\frac{0.810}{t_{e}}-0.502log(t_{e}), 
\label{equ_6}
\end{eqnarray}

\begin{eqnarray}
12+log\left(\frac{Ar^{3+}}{H^{+}}\right)=log\left(\frac{I(4740)}{H_{\beta}}\right)+6.362 + \nonumber\\
+ \frac{1.174}{t_{e}}-0.820log(t_{e}),
\label{equ_7}
\end{eqnarray}

\noindent where we assumed $t_{\rm e}$ = $t_{\rm e}$[\ion{Ar}{iii}] $\simeq$ ~$t_{\rm e}$[\ion{S}{iii}] in the first expression (Eq.~(\ref{equ_6})) and $t_{\rm e}$ = $t_{\rm e}$[\ion{Ar}{iv}] $\simeq$ ~$t_{\rm e}$[\ion{O}{iii}] in the second ( Eq.~(\ref{equ_7})). All of them are in units of 10$^{4}$K.

When a spectrum does not clearly show the argon lines, the contribution by S$^{3+}$ is not necessarily negligible because these are weak lines, and spectra with a very high signal to noise ratio are required to measure these lines reliably. However, the presence or absence of S$^{3+}$ depends directly on the hardness of the ionizing radiation, since high-energy photons are required to ionize sulphur three times. \cite{vil88} defined the $\eta$ parameter, which they called the{\em softness parameter}. This parameter is inversely proportional to the radiation hardness,  

\begin{equation}
\eta = \frac{O^{+}/O^{++}}{S^{+}/S^{++}}.
\label{eq_eta}
\end{equation}

\begin{figure}
\includegraphics[width=0.4\textwidth]{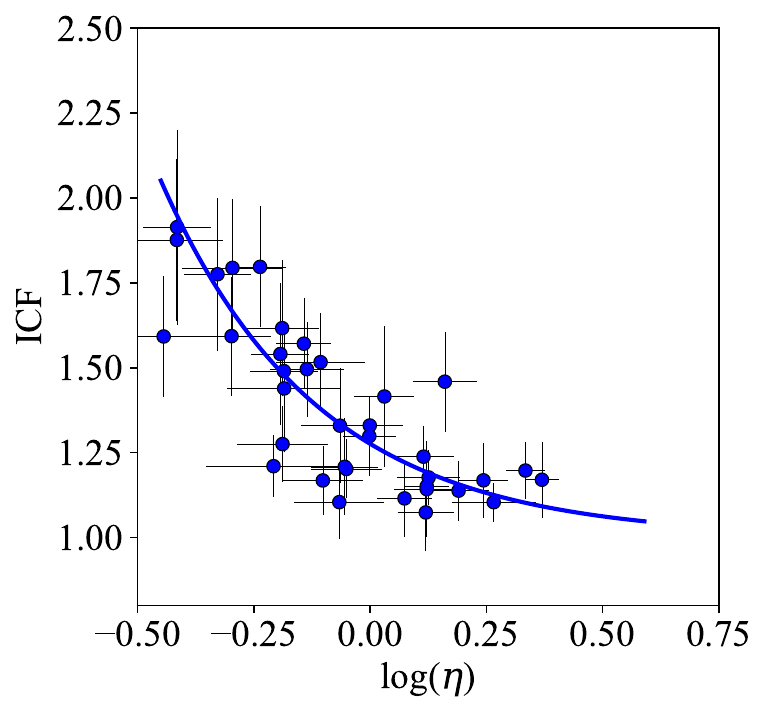}
\caption{Relation of ICF and the hardness of the radiation parameterized by $\eta$. Objects with log($\eta) \geq$ 0.5 are assumed to have ICF = 1. The blue line shows the fit defined by Eq.~(\ref{ecuacion_eta}).} 
\label{eta}
\end{figure}

Following DZ22,  we represent in Fig.~\ref{eta} the computed ICF for the objects in sample 3 with argon lines as a function of log($\eta$). The oxygen ionic abundances were calculated as described in Sect.~\ref{oxigeno}. The function that fits the data best is 

\begin{equation}
log(ICF-1) = (-1.225 \pm 0.057) + (-0.561 \pm 0.018) \cdot log(\eta).
\label{ecuacion_eta}
\end{equation}

This relation allowed us to calculate the ICF for objects without the [\ion{Ar}{iv}] line, but with a possibly substantial contribution by S$^{3+}$. According to Fig. \ref{eta} and Eq.~(\ref{ecuacion_eta}), for objects with log($\eta$) > 0.5,  ICF(S) = 1. This means that the radiation is not hard enough to ionize sulphur three times. 

Following this procedure, we calculated the total sulphur abundances for the objects in sample~3. The total and ionic abundances for the objects of this sample are given in Table~2.

Because the auroral line [\ion{S}{iii}]$\lambda$6312 \AA\ ~is detected in all objects of sample 2, we calculated the $\frac{\left(S^{+}+S^{++}\right)}{H^+}$ abundance with the direct method, although the ICF cannot be estimated in this case. The corresponding abundances are shown in electronic form at the CDS.

\subsection{Empirical calculation of the sulphur abundance}
\label{empirical}

When the spectra do not present all the lines required to calculate the electron temperature directly, the preferred option is to determine the metallicity abundances using intense lines through an empirical relation with some adequate parameter. In the case of sulphur, the chosen parameter for calibrating the abundance is S$_{23}$

\begin{equation}
    S_{23} = \frac{I(6717) + I(6731) + I(9069) + I(9532)}{I(H_{\alpha})}\cdot \frac{I(H_{\alpha})}{I(H_{\beta})}.
\end{equation}

\noindent We followed the calibration obtained by DZ22, which is valid for metallicities up to solar,

\begin{eqnarray}
\label{calibrador_DZ22}
12+log(S/H) = (1.060 \pm 0.098) \: log^{2}(S_{23}) + \nonumber\\ (2.202 \pm 0.050) \: log(S_{23}) + (6.636 \pm 0.011).
\end{eqnarray}

Based on this relation, we were able to calculate the total sulphur abundance for the objects in the complete sample (sample~1).


\section{Discussion}
\label{discusion}


\begin{figure}
\includegraphics[width=0.5\textwidth]{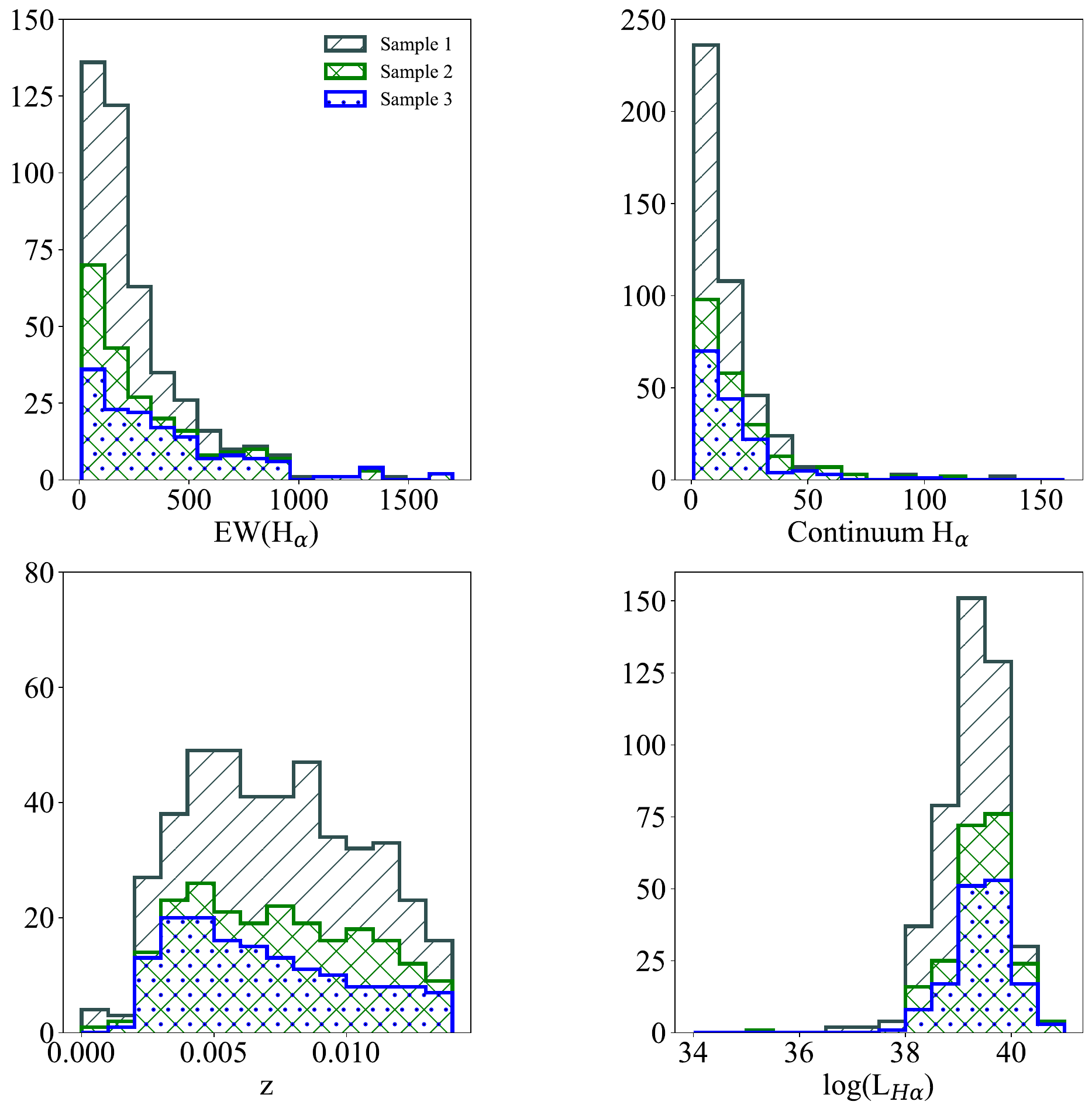}
\caption{Characterization of the total sample and of the respective subsamples.}
\label{caracter}
\end{figure}

\subsection{Physical conditions}
Figure \ref{caracter} shows the characterization of the sample and highlights the constraints we imposed to select it.

\begin{figure}
\includegraphics[width=0.5\textwidth]{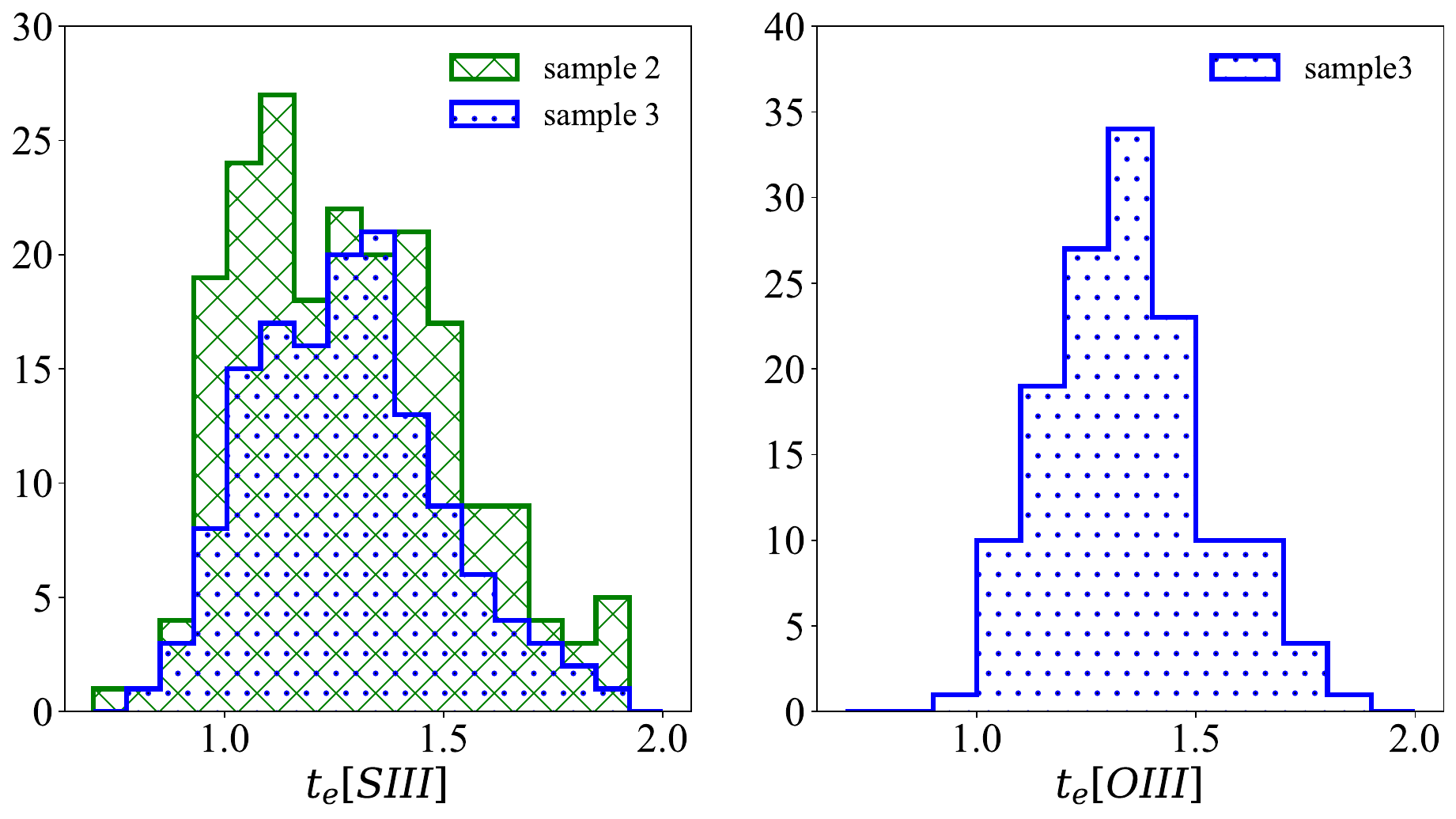}
\caption{Electron temperature distribution. Left panel: Sulphur electron temperature for objects in samples 2 and 3. Right panel: Oxygen electron temperature for sample 3. Objects in sample 2 do not show the \OIII{4363} auroral line, and therefore, no $t_{\rm e}$[\ion{O}{iii}] is available for these objects.}
\label{histo_te}
\end{figure}

In the left panel of Fig. \ref{histo_te}, we present the distribution of $t_{\rm e}$[\ion{S}{iii}] for the objects in samples 2 and 3. As a result of the conditions we imposed in the selection of the samples, the electron temperature is biased toward high values. Objects with lower electron temperatures than the lower limit shown by the distribution belong to sample~1. The distribution of $t_{\rm e}$[\ion{O}{iii}] is shown in the right panel of the figure. The bias introduced when the \OIII{4363} \AA\ line is required is reflected in the fact that objects with electron temperatures lower than ~10,000K are absent from our sample. This causes the temperature distribution to be somewhat narrower. This concentration around a central value is absent in the case of $t_{\rm e}$[\ion{S}{iii}], where the distribution is flatter and wider. This fits a more realistic scenario better.

\begin{figure}
\includegraphics[width=\linewidth]{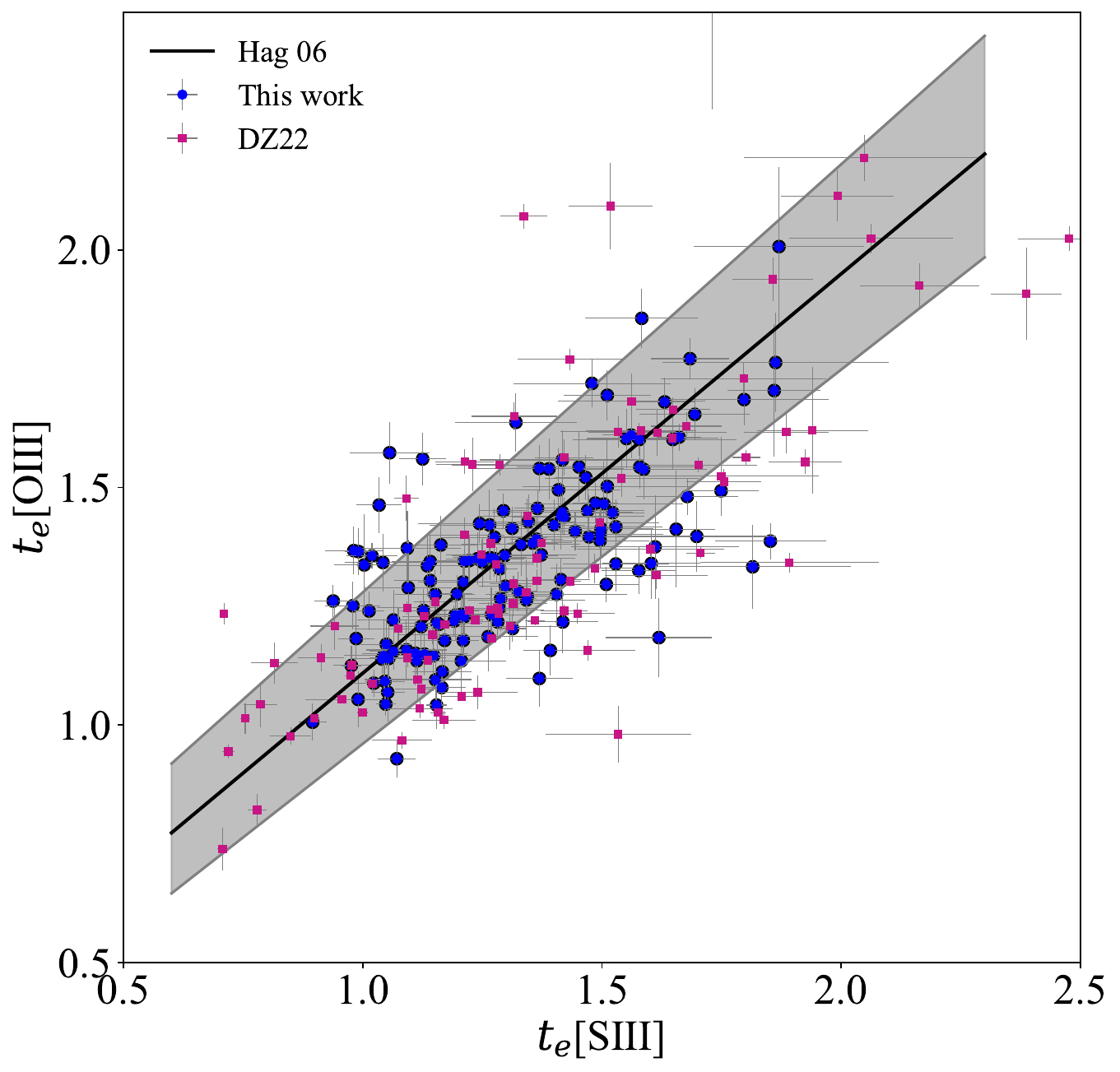}
\caption{Relation of the oxygen and sulphur electron temperatures. The blue circles show the values for objects in sample~3, and magenta squares show data from DZ22 and references therein. The black line shows the fit from \cite{hag06}}
\label{teO_teS}
\end{figure}

Figure~\ref{teO_teS} shows the comparison between [\ion{O}{iii}] and [\ion{S}{iii}] electron temperatures. The solid black line shows the fit to the empirical data from \cite{hag06}, and the shaded area delimits the estimated error band in the fit. Most of the objects analyzed here fall inside this area. 

\begin{figure}
\centering
\includegraphics[width=0.40\textwidth]{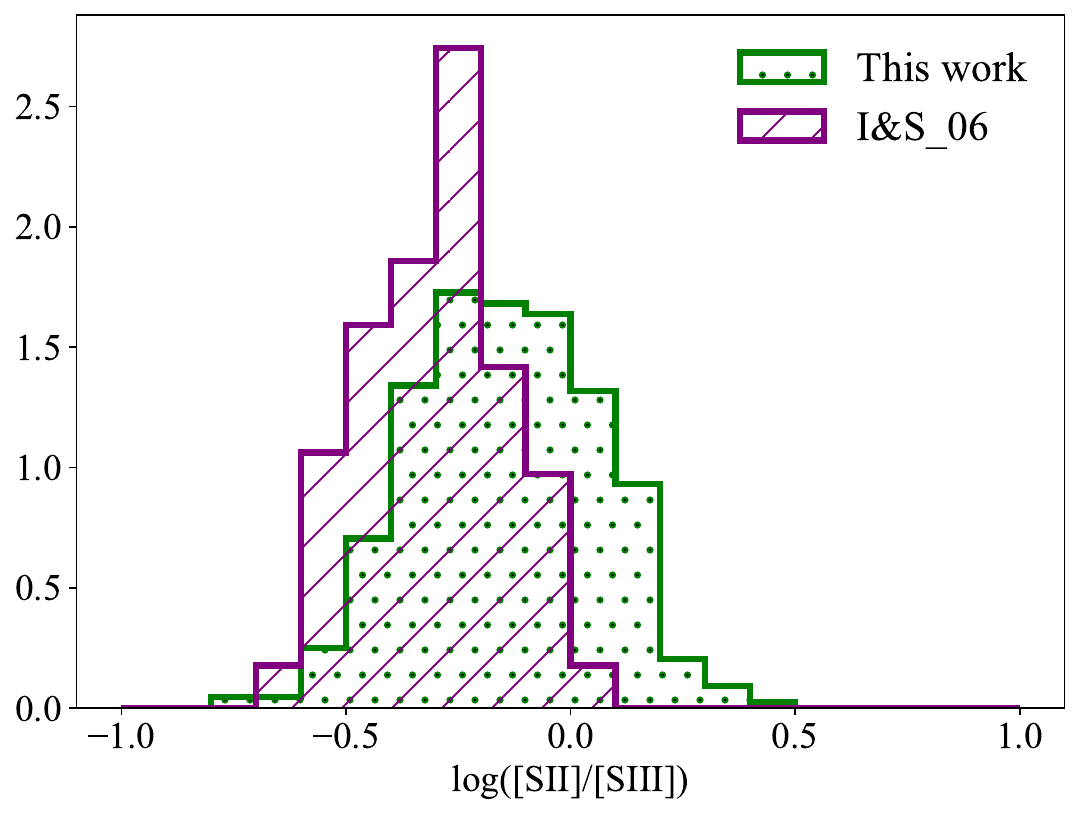}
\caption{Distribution of the ionization parameter traced by the ratio of the sulphur emission lines for objects in sample 2 in green, compared with the results from \cite{izo06} in magenta.}
\label{histo_u}
\end{figure}

One of the physical properties of the gas that can give information about the interaction between the ionization source and the ionized gas is its degree of ionization. It is often characterized by the ionization parameter, u. Defined as the ratio of the ionizing photon flux and the gas electron density, it is traced well by the emission-line ratio of \SII{6717} + \SII{6731}~\AA  \, and \SIII{9069} + \SIII{9532} \AA\ \citep{dia91}. We estimated this value for our whole sample and present its distribution in Fig.~\ref{histo_u}, where we compare it to the derivation obtained from the data in \cite[SDSS, DR3]{izo06}. The former looks wider because the requirement of the detection of the \OIII{4363} line in the latter work introduces an important bias that selected the objects with a higher ionization parameter \citep{hoy06}.

The distribution we found presents higher values on average than were obtained for HII regions in galaxies (see DZ22). This indicates a greater number of ionizing photons and less evolved ionizing stellar populations.

\subsection{Ionic abundances}

The main assumption underlying the derivation of the sulphur ionic abundance is that the [\ion{S}{iii}] electron temperature is representative of an extended zone in which the low- and intermediate-ionization zones overlap and in which a common electron temperature is therefore assumed. 
Today, empirical evidence of a tight relation that is very close to one of $T_{\rm e}[\ion{N}{ii}]$ and $T_{\rm e}[\ion{S}{iii}]$ abounds across a wide range of electron temperatures
\citep{dia07,cro16,ber20,rog21,ric24}. 
On the other hand, measurements of $T_{\rm e}[\ion{O}{ii}]$, $T_{\rm e}[\ion{S}{ii}]$, and $T_{\rm e}[\ion{N}{ii}]$ indeed suggest that both ions  arise from a common ionization zone \citep{cro16}, and as expected for two ions that probe similar low-ionization gas, the best fit for their electron temperatures is consistent with equality \citep{ber20}. It is therefore reasonable to assume that $T_{\rm e}[\ion{S}{iii}]$ and $T_{\rm e}[\ion{S}{ii}]$ have similar values. However, \citet{rog21} reported that the $T_{\rm e}[\ion{S}{ii}]$ versus $T_{\rm e}[\ion{N}{ii}]$ relation is not entirely consistent with a slope of unity. The general trend is offset toward higher $T_{\rm e}[\ion{S}{ii}]$ at fixed $T_{\rm e}[\ion{N}{ii}]$, although the errors are large and there are some deviating outliers.
Interestingly enough, the samples analyzed by \citet{ber20} and  \citet{rog21} only differ in the added 24 regions of NGC~2403, and in 17 of these 24 regions, the values of  $T_{\rm e}[\ion{S}{ii}]$  are equal to those of $T_{\rm e}[\ion{S}{iii}]$  within the observational errors.

The derived relation in \citet{rog21} yields values of $T_{\rm e}[\ion{S}{ii}]$ that are equal to $T_{\rm e}[\ion{N}{ii}]$ within the observational errors for [\ion{N}{ii}] electron temperatures lower than 12000 K. 
If $T_{\rm e}[\ion{S}{ii}]$ is indeed higher than $T_{\rm e}[\ion{N}{ii}]$, our S$^+$ abundance would be overestimated, as would its contribution to the total sulphur abundances. For a $T_{\rm e}[\ion{S}{ii}]$ higher than $T_{\rm e}[\ion{S}{iii}]$ $\sim$ $T_{\rm e}[\ion{N}{ii}]$, the derived S$^{+}$/H$^{+}$ ionic abundance would be lower. On the other hand, however, the nebula would be more strongly dominated by S$^{++}$, and the resulting total S/H would be almost unchanged because the contribution of S$^{+}$ to the total abundance would be very low. For example, for an object with $T_{\rm e}[\ion{S}{iii}]$ $\sim$ 20000K, an increase in $T_{\rm e}[\ion{S}{ii}]$ by a factor of two  decreases  the S$^{+}$/H$^{+}$ abundance by a factor of about 2.4 (0.38 dex) while the total (S$^{+}$/H$^{+}$) + (S$^{++}$/H$^{+}$) changes by a factor of only 1.15 (0.06 dex) because the S$^{+}$/S$^{++}$ ratio is only 0.12. Therefore, the assumption of $T_{\rm e}[\ion{S}{ii}]$ $\sim$ $T_{\rm e}[\ion{S}{iii}]$ can be made in most cases without introducing large errors.

In ionized regions with a low to moderate metal abundance, it is commonly assumed that most of the sulphur is in the form of once- and twice-ionized species. However, in a number of the HII galaxies studied here, the temperature of the ionizing source is high enough for some contribution by S$^{3+}$to be expected. Since no emission lines, mainly the [\ion{S}{iv}] line at 10.51 $\mu$, are present in the available spectral range to derive this ionic abundance, we needed to calculate an ionization correction factor (ICF) to estimate its contribution to the total sulphur abundance.

The ICF of sulphur is typically calculated from the relation of the O$^{+}$ and O$^{++}$ ionic abundances  \citep[see][]{pei69}. Many authors investigated possible ways of calculating the ICF with the use of photoionization models \citep[][]{sta78,bar80}. Based on the photoionization models of \cite{sta78}, the following fitting function can be written:

\begin{equation}
ICF = \left[ 1-\left( 1-\frac{O^{+}}{O}\right) ^{\alpha}\right] ^{-1/\alpha}.
\label{stasinska}
\end{equation}

 The value of the $\alpha$ exponent is found to be between 2 and 3.5. \cite{dor16} placed it at 3.27. This method relies on the knowledge of well-determined abundances of O$^{+}$, and hence, $t_{\rm e}$[\ion{O}{ii}], which is a problem that remains to be solved.

\begin{figure}
\includegraphics[width=0.5\textwidth]{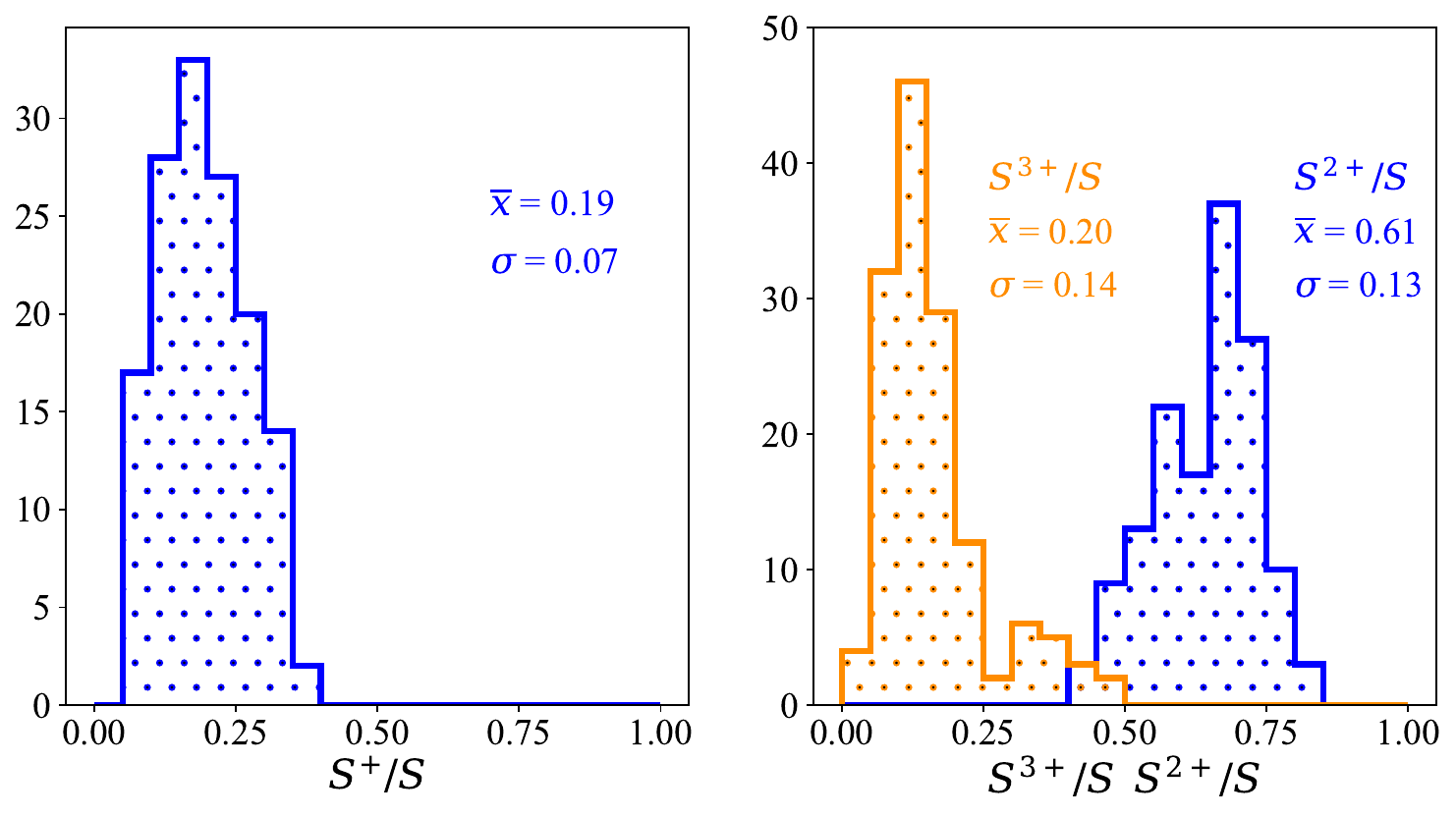}
\caption{Ionic ratio distributions. The left panel shows the relation of S$^{+}$ and the total sulphur abundance, and the right panel presents the same for S$^{2+}$ (blue) and S$^{3+}$ (orange).} 
\label{histo_proporciones_ionicas}
\end{figure}

This way of calculating the ICF is a consequence of choosing oxygen as the main element to trace the metallicity. Other ways to obtain the ICF independently of the oxygen abundance are available today. As mentioned above, \cite{vit18} calculated the sulphur ICF from the ratio of two consecutive ionization ratios of sulphur and argon (see Sect. \ref{S-ICF}).
The application of this method requires the detection of the argon lines \ArIII{7135} \AA\ and \ArIV{4740} \AA\,, the latter of which is rather weak. Unfortunately, this is therefore not always possible. However, it seems reasonable to assume that the ionization of S$^{++}$
requires the production of high-energy photons from the ionizing source. This assumption leads to the possible existence of an anticorrelation between the $\eta$ softness parameter and the ICF (see Eq.~(\ref{ecuacion_eta}) and Fig. ~\ref{eta}).
For objects that lack argon lines and oxygen ionic abundances, we estimated the value of $\eta$ with the use of this equation. This was only possible for 24 \% of the objects in sample~3. Figure \ref{eta} shows that only objects with values of log($\eta$) < 0.4 \citep[i.e., effective temperature of the ionizing radiation of about 40000 K in Mihalas' scale][]{vil88} require the application of an ICF. For values of log($\eta$) greater than 0.5, the radiation is not hard enough to ionize sulphur three times, and in these cases, the ICF was set equal to 1. 
Following this procedure, we calculated the abundances of the sulphur ions and their ratios. The distributions are shown in Fig.~\ref{histo_proporciones_ionicas}.  S$^{2+}$ is found to be the dominant ion, with a contribution greatly exceeding that of S$^{+}$, as expected. This characteristic means that the large uncertainties  in $t_{\rm e}$[\ion{S}{ii}] are not significant in any case, because S$^{+}$/S$^{++}$ < 1 even in objects with the lowest excitation. For instance, for the low-excitation region CCM10 in M51, O$^{+}$/O$^{++}$ $\sim$ 6, and S$^{+}$/S$^{++}$ $\sim$ 0.3 \citep{bre04}. 
This fact justifies considering  T$_{\rm e}$[\ion{S}{iii}] as more representative of the average T$_{\rm e}$ in the nebula than either T$_{\rm e}$[\ion{O}{ii}] or T$_{\rm e}$[\ion{O}{iii}]


The right  panel of the figure shows that in a non-negligible number of objects, 30 out of 141, the contribution of S$^{3+}$ to the total sulphur abundance is equal to or greater than 20\%. This fact highlights the importance of the ionization correction factor. The ionic abundances of sulphur and argon are presented in Table~2.

\subsection{Total abundances}

\subsubsection{Direct abundances}
\label{direct}
\begin{figure}
\includegraphics[width=0.5\textwidth]{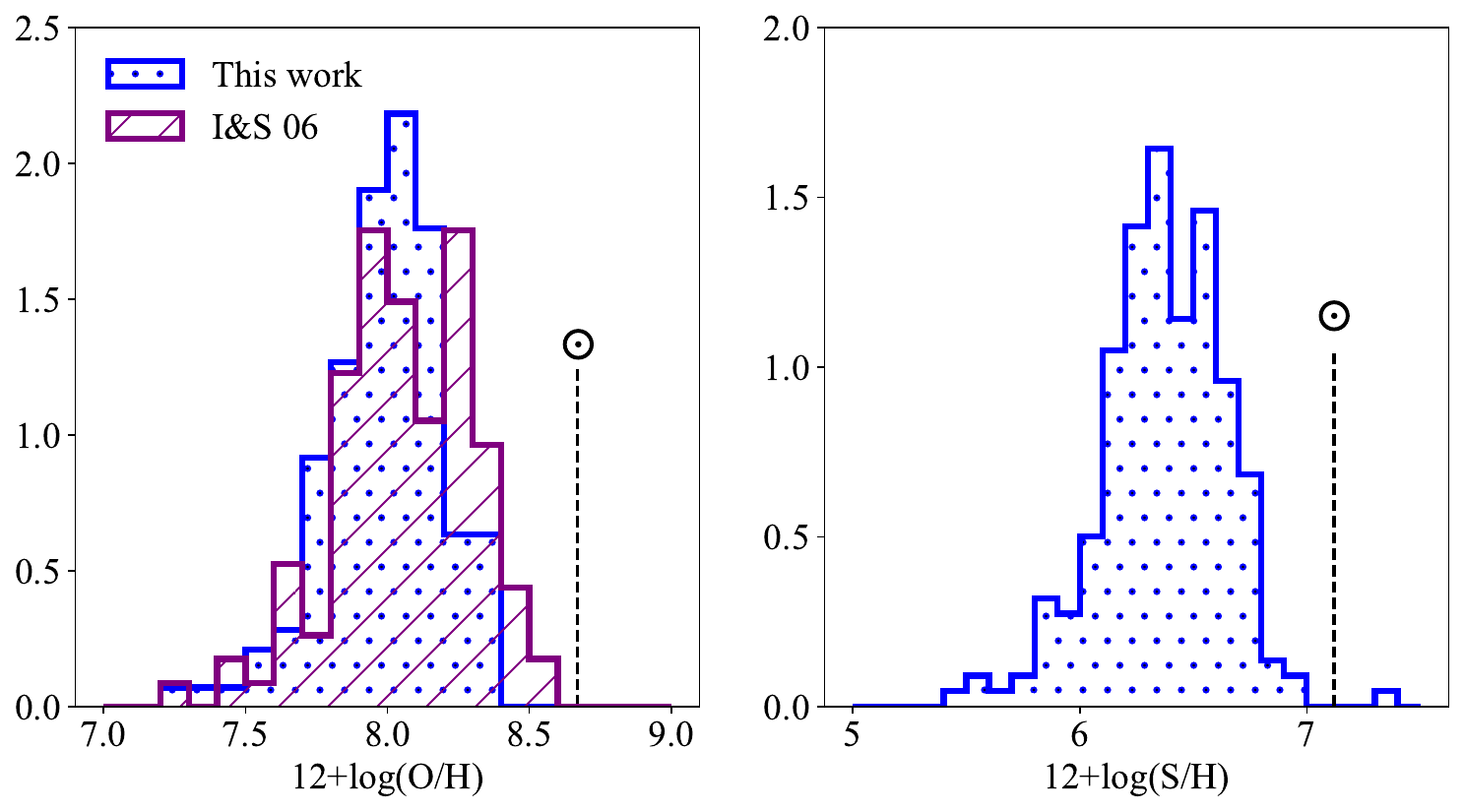}
\caption{Left panel: Directly derived oxygen abundance distribution for objects in sample~3 compared with the results from \cite{izo06}. Right panel: Directly derived sulphur abundance distribution for objects in sample 2.} 
\label{histo_OH}
\end{figure}

The direct abundance of oxygen can only be calculated for objects in which the \OIII{4363} line is detected. These objects are those in sample 3. As mentioned above, this constraint introduces a bias toward objects with high values of $t_{\rm e}$[\ion{O}{iii}] and hence low metallicities, which never reach the solar value. We compared our results, obtained with the temperature scheme described above, with those obtained by \cite{izo06}, who adopted a different temperature structure. The result is shown in the left panel of Fig.~\ref{histo_OH}. The results are clearly very similar, with both distributions corresponding to objects in the same metallicity range. The reason is that by imposing the same restriction on the presence of the \OIII{4363} line in both cases, the gas excitation is high, and therefore, the contribution of O$^{+}$ to the total oxygen abundance is  small (for only 13 objects in our sample, O$^{+}$ is comparable to O$^{++}$). Under these conditions, the adoption of different assumptions for the ionization structure, which mainly affects the derivation of the [\ion{O}{ii}] electron temperature, is almost irrelevant. 

The total abundances of sulphur for the objects in sample 2 are shown in the right  panel of Fig~\ref{histo_OH}. In this case, 38\% of the objects have no ICF correction. These objects belong to sample~2, but do not overlap with sample~3 because they are those showing the \SIII{6312} line, but not the \OIII{4363} line. This indicates that their electron temperature is low. Under these conditions, their contribution  by S$^{3+}$ to the total sulphur abundance is expected to be low, and their ICF should be close to unity. 

\subsubsection{Calibrated abundances}

The derivation of the sulphur abundances for sample~1 objects which lack the 6312 \AA\ emission line required the use of empirical methods. The S$_{23}$ calibrator, described in Sect. \ref{empirical}, is based on the nebular lines of sulphur. Its advantage is its linear behavior over a wide range of metallicities that reaches the solar value, contrary to the case of the widely used R$_{23}$, which uses strong oxygen emission lines. The nebular sulphur lines in the far red are easily detected in the range of moderate to high abundances \citep[see][]{dia91}. Only in cases of very low metallicities in which the absolute value of the sulphur abundance is very low, the cooling is not dominated by collisionally excited lines of either sulphur and/or oxygen. Fortunately, this is not the case for most HII galaxies, whose metallicity, although low, is higher than this minimum value.

Assuming the hypotheses stated by \cite{pag79} for the validity of the calibrations using intense lines, there must be a relation of the electron temperature and the abundance of the element under study. On the other hand, the dependence of the calibrator on the ionization parameter is expected to be very weak.

\begin{figure}
\begin{center}
\includegraphics[width=0.38\textwidth]{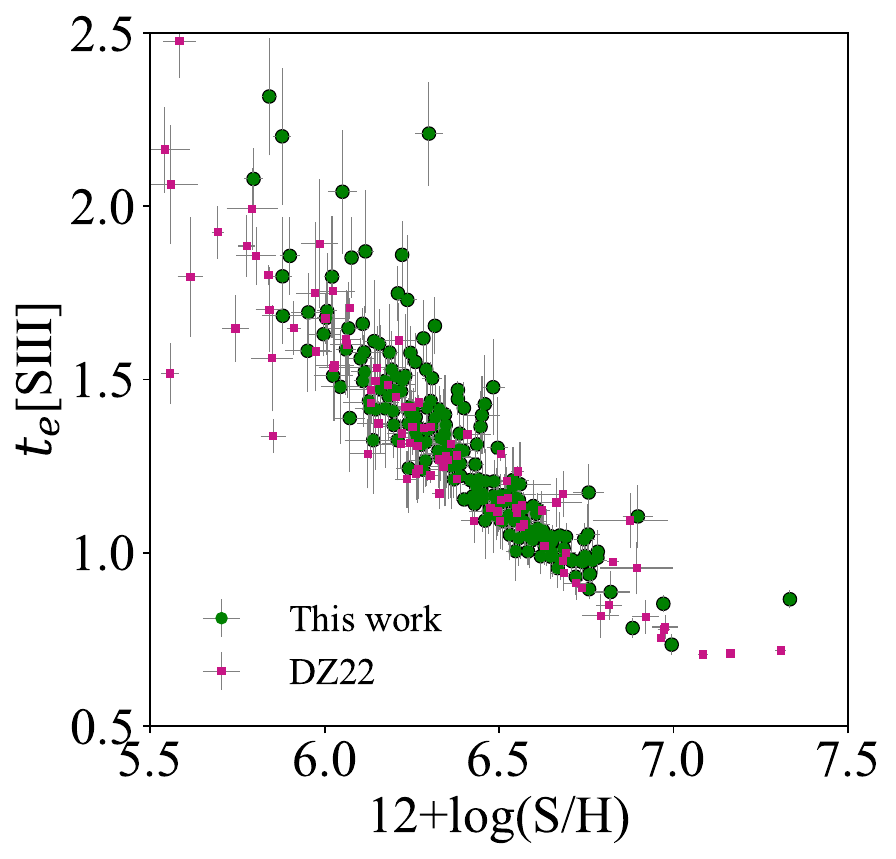}
\caption{Relation of the sulphur electron temperature and the total sulphur abundance for the objects in sample~2. The magenta squares correspond to the data compilation by DZ22.} 
\label{teS_SH}
\end{center}
\end{figure}

In Fig. \ref{teS_SH} this relation is shown for the objects of this study and for the observations described in DZ22 (which cover a wider range of metallicities) for the same type of object. They agree very well. This allowed us to use the function obtained by DZ22, that is, Eq.~ (\ref{calibrador_DZ22}), to fit the calibration of the S$_{23}$ parameter for the objects of sample 1.

\begin{figure}
\begin{center}
\includegraphics[width=0.5\textwidth]{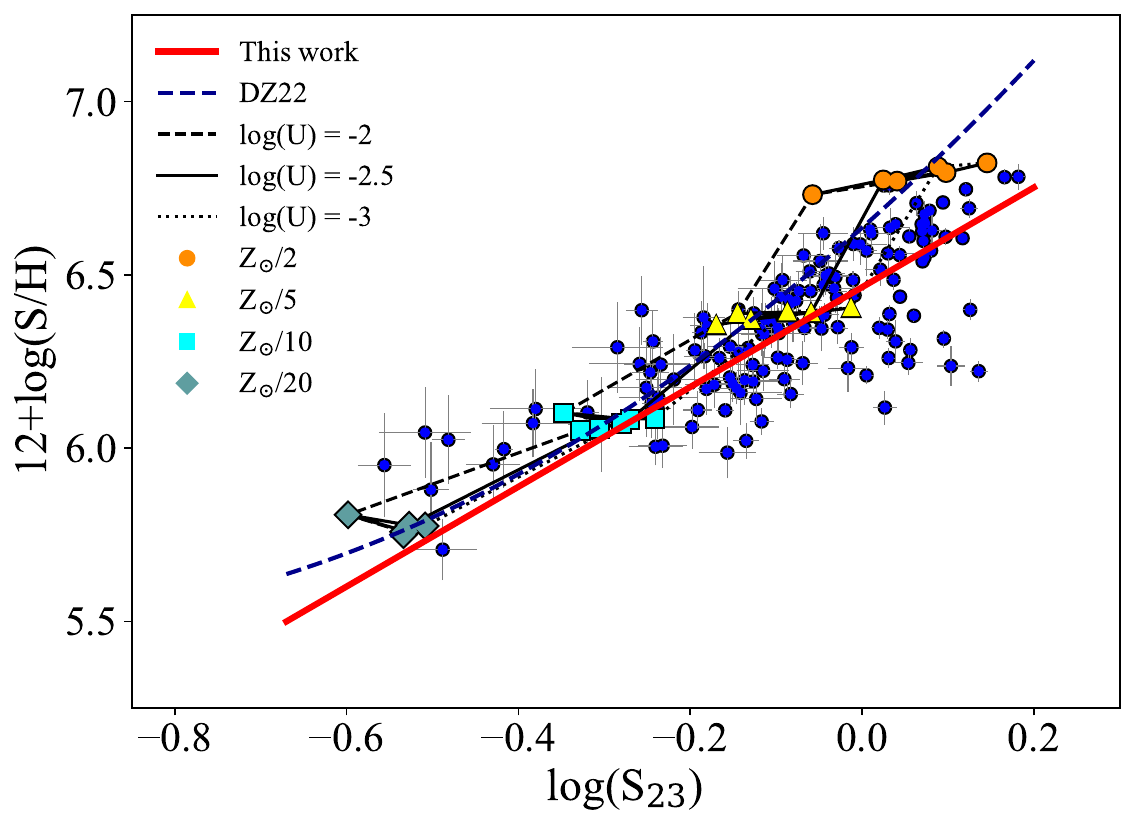}
\caption{Relation of the total sulphur abundance and the S$_{23}$ parameter. The blue circles correspond to sample~3 data, the large filled symbols are those from Cloudy models, and the black lines join models with the same ionization parameter. The dashed blue line shows the relation by DZ22, and the solid red line shows the quadratic fit of the data.} 
\label{S23}
\end{center}
\end{figure}

In Fig.~\ref{S23} we show the actual calibration of the S$_{23}$ parameter using only the data analyzed in this work. The calibration by DZ22 has been presented for comparison purposes, and it shows good agreement with the best quadratic fit to our data. In the high-metallicity range, the two fittings diverge. One fit corresponds to our data and is below the one we applied to the objects in sample 1. Again, this is a selection effect that biases the sample selection toward objects with a low metallicity and high temperature. The upper right area of the plot is expected to be populated by objects if the sample metallicity range were wider, as is the case for the sample of DZ22. Therefore, if the calibration were only restricted to the objects studied here, the identification of the HII galaxies with the highest metallicities would not be possible. As previously mentioned, the calibration is single-value up to solar metallicity, a range that includes most of the HII regions and HII galaxies. This is in contrast to  O$_{23}$\footnote{The sum of the [\ion{O}{ii}]}$\lambda \lambda 3727,3729$ \AA\ and [\ion{O}{iii}]$\lambda \lambda 4959,5007$ \AA\ relative to the H$_{\beta}$ lines, which becomes twofold for metallicities higher than 12 + log(O/H) $\sim$ 8.2.

 We also show in Fig. \ref{S23}  the results of the Cloudy models described in Sect. \ref{S-ICF}. The different colored filled symbols correspond to different metallicities, as labeled, from 5 to 50\% of the solar value, and are joined by lines of the same ionization parameter, u. These lines are very close together and sho- the low ionization parameter dependence of \ref{S23}. On the other hand, the different metallicity symbols, grouped by ionization parameter, are well separated along the calibration line. 

\subsubsection{HII galaxy abundance distribution}

\begin{figure}
\begin{center}
\includegraphics[width=0.5\textwidth]{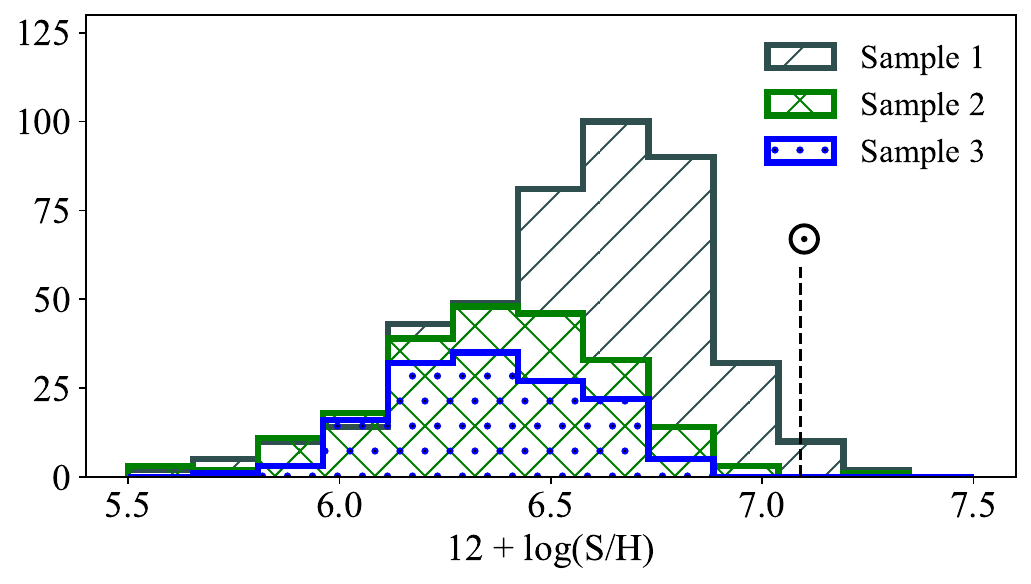}
\caption{Total sulphur abundance distribution. The dashed black line corresponds to the solar value. The values are not normalized to reveal the difference in the number of objects in each sample. } 
\label{SH_todas}
\end{center}
\end{figure}

Figure \ref{SH_todas} shows the abundance distribution of the three samples. With relaxing restrictions on the presence of auroral lines, the abundance of the sample extends farther toward higher values. Only objects of sample 1 reach the sulphur solar abundance, and a few of them present even higher values than this.
The distributions of samples 2 and 3 are very similar to each other and peak at a the same sulphur abundance because both were derived using direct methods that require the detection of weak lines, and therefore, high signal-to-noise ratios. We recall that the nondetection of electron-temperature-sensitive lines does not necessary imply a low metal content, but it could be due to the presence of a stronger underlying stellar continuum or a noisy spectrum \citep{hoy06}. On the other hand, any uncertainty attributed to the assumed ionization structure ($T_{\rm e}[\ion{S}{ii}]$ equal to $T_{\rm e}[\ion{S}{iii}]$) has very little effect on the derivation of the sulphur abundance. As pointed out above, the contribution of S~$^{+}$ to the total abundance is small for most of our objects (see Fig.~\ref{histo_proporciones_ionicas}). 

The abundance distribution in HII galaxies has rarely been studied using homogeneous samples. The first comparison sample we found is the one referenced in Sect. \ref{direct} by \citet{izo06}, for which we derived the oxygen abundance distribution shown in Fig. \ref{histo_OH}. The distribution shows two peaks at 12+log(O/H) with values of 8.0 and 8.4 and a median of 8.1 (22\% of the solar value). This is exactly the same median value as we obtained for our distribution, although we assumed different ionization structures. However, in the case of sulphur, the median is slightly lower than the solar value: [12+log(S/H)]$_{median}$  $\simeq$ 6.37 (18 \% solar). In all cases, the abundances were derived using the direct method. These values yield an S/O ratio of [log(S/O)]$_{median}$= -1.7, which is close to the solar value (-1.5).

The second study that we found was by \citet{shi10}. These authors selected a sample of 5880 star-forming galaxies from the SDSS DR4 and provided two oxygen-abundance distributions. The first abundance was derived using a model-based electron temperature following \citet{pil01}, and the second abundance was derived using a Bayesian approach. While the former showed a narrow symmetric profile concentrated around a high peak at 12+log(O/H) about 8.35, the latter showed a wider asymmetric shape that resembles the shape we found in our Fig. \ref{SH_todas}. Since the Bayesian method is based on simultaneous fits of all the most prominent emission lines and not only on those of the oxygen species \citep[see also][]{vit19}, this may reflect the fact that this approach adds complementary information at longer wavelengths that are similar to the wavelength provided by the use of the sulphur lines.

\section{Conclusions}
\label{conclusiones}

We have measured the fluxes of the main emission lines in the spectra of a selection of HII galaxies from the 16th release of SDSS, including the [\ion{S}{iii}] $\lambda$ 9069 \AA\ line. This selection provided an homogeneous sample of 439 objects with a redshift of z $\leq$ 0.014. In order to further maintain this homogeneity, we developed our own tool to perform the measurements. This produced slightly higher values than were provided by the SDSS pipeline. Out of the main sample of 439 objects, the [\ion{S}{iii}]  $\lambda$ 6312 \AA\ auroral line was soundly detected in 224 objects. This allowes us to derive the sulphur abundances with the so-called direct method, following the method presented in \citet{dia22}. When the auroral line was not detected, we used the strong-line empirical method by means of the S$_{23}$ abundance parameter following the calibration of \cite{dia22} to estimate the sulphur abundance. 

For the objects that required the detection of the \OIII{4363}~\AA\ line (sample~3) and [\ion{S}{iii}]  $\lambda$ 6312 \AA\ (sample~2), the directly derived distribution abundances as traced by oxygen and sulphur (see Fig. \ref{histo_OH}) appear to be very similar to each other. The median values are 12+log(O/H) = 8.1 and 12+log(S/H) = 6.4, which corresponds to an S/O ratio of log(S/O)= -1.7 and is close to the solar value (-1.5). However, when we relaxed the restriction of lines that are sensitive to weak temperatures, the abundance distribution was wider, with a median value of 12+log(S/H) = 6.6. We recall that this latter distribution included objects with abundances that were derived from the empirical calibration of sulphur, which is much more reliable than the calibrations used for the strong oxygen lines. The sulphur calibrator $S_{23}$ decreases monotonically up to supersolar values, while the widely used $R_{23}$ oxygen calibrator is two-fold and reverses at the oxygen abundance value at which more than 50\% of the HII galaxies are located (see \citep[see][]{per05}). If the S/O ratio were assumed to be constant, the median sulphur abundance value found here would imply an oxygen abundance median value of 12+log(O/H) = 8.3. 

In summary, the abundance distributions traced by sulphur can reach reliable abundances up to the solar value at least. They provide a more complete picture of the metallicity distribution of HII galaxies.
A further strength of the method is that its application only involves the red part of the spectrum (between 6000 and 9600 $\AA$), where the effect of reddening is low. Although the strong nebular [\ion{S}{iii}] lines shift beyond the far red spectral region for high-redshift objects, present-day infrared spectrographs can overcome this difficulty, and observations made with NIRSpec on board the JWST are expected to be able to provide data for objects with redshifts between 0 and 4.24.

 \section*{Data availability}
Full Table 1, Table 2 as well as Table 3 presenting the references for the studied objects, are provided only in electronic form at the CDS via anonymous ftp to cdsarc.u-strasbg.fr (130.79.128.5) or via http://cdsweb.u-strasbg.fr/cgi-bin/qcat?J/A+A/.

\begin{acknowledgements}
This work is based on data from SDSS.  Funding for the Sloan Digital Sky Survey IV has been provided by the Alfred P. Sloan Foundation, the U.S. Department of Energy Office of Science, and the Participating Institutions. SDSS-IV acknowledges support and resources from the Center for High-Performance Computing at
the University of Utah. The SDSS web site is www.sdss.org.

The STARLIGHT project is supported by the Brazilian agencies CNPq, CAPES and
FAPESP and by the France-Brazil CAPES/Cofecub program.

This research has been funded through grants  PID2019-107408GB-C42 and PID2022-136598NB-C33 by MCIN/AEI 10.13039/501100011033 and  by “ERDF A way of making Europe”.

\end{acknowledgements}







\end{document}